\documentclass[aps,prb,reprint,groupedaddress]{revtex4-2}
\usepackage{graphicx}
\usepackage{amsmath}
\usepackage{multirow}
\begin{document}


\title{Delaying two-photon Fock-states in a hot cesium vapor using on-demand generated single-photons from a semiconductor quantum dot}


\author{H.~Vural$^{1}$}
 

\author{S.~Seyfferle$^{1}$} 


\author{I.~Gerhardt$^{2}$} 

\author{M.~Jetter$^{1}$} 

\author{S.~L.~Portalupi$^{1}$} 

\author{P.~Michler$^{1}$}

\email[]{p.michler@ihfg.uni-stuttgart.de}
\affiliation{$^1$ Institut f\"ur Halbleiteroptik und Funktionelle Grenzfl\"achen, Center for Integrated Quantum Science and Technology ($IQ^{ST}$) and SCoPE, University of Stuttgart, Allmandring 3, 70569 Stuttgart, Germany\\
$^2$ 3. Physikalisches Institut and Center for Integrated Quantum Science and Technology, University of Stuttgart, Pfaffenwaldring 57, D-70569 Stuttgart, Germany
}


\date{\today}

\begin{abstract}
Single photons from solid-state quantum emitters are playing a crucial role in the development of photonic quantum technologies. Higher order states, such as $N$-photon Fock-states allow for applications e.g. in quantum-enhanced sensing. In this study, we utilize the dispersion of a hot cesium vapor at the D$_1$ lines to realize a temporal delay for one and two-photon Fock-states as a result of the slow-light effect. Single photons are generated on-demand from an InGaAs quantum dot, while their quantum interference at a beam splitter is used to generate a two-photon Fock-state. We verify the successful propagation and temporal delay of both the one and two-photon Fock-states, while a significant delay (5 $\times$ initial photon length) with simultaneous high transmission ($\sim90\,$\%) is achieved.
\end{abstract}


\maketitle

\section{Introduction}
Self-assembled semiconductor quantum dots (QD) are one of the appealing platforms for the realization of optical quantum technologies~\cite{Michler:2017}. Under resonant $\pi-$pulse excitation QDs are promising on-demand emitters of coherent, indistinguishable single~\cite{He.He.ea:2013} and entangled photons~\cite{Muller.Bounouar.ea:2014,Liu.Su.ea:2019}. The ultra high rates achieved by integration of QDs in photonic cavity structures pushed the frontier of this photonic technology to a level competitive with computer simulations in special tasks like boson sampling~\cite{Wang.Qin.ea:2019}.

A recently evolving field is the generation of two-photon states. These states are highly beneficial for quantum technology applications like quantum metrology~\cite{Nagata2007,Mitchell2004,Mueller.Vural.ea:2017}, quantum key distribution~\cite{Jennewein2000}, as well as for the implementation of quantum repeaters~\cite{Simon2007} and even in fields such as two-photon microscopy~\cite{Upton2013}.

\noindent Proposals for efficient two-photon generation from semiconductor QDs include cavity-enhanced recombination~\cite{Valle2011,Munoz2014}, with the added benefit of bypassing the finestructure splitting of the exciton states and thereby preserving a high degree of entanglement~\cite{Schumacher2012}. In recent experimental demonstrations, two-photon states have been identified in 
the superradiant emission of two quantum dots coupled to the same waveguide mode~\cite{Kim2018} and in the emission of a single QD two-level system under resonant pumping at a $2\pi$ pulse, while the system's re-excitation during a single excitation pulse yields contributions with multi photons~\cite{Fischer2017,Loredo2019}. The coherence of the multi-photon contributions and their probability amplitudes have been investigated which strongly depend on excitation pulse length and area. The $\pi$ pulse excitation leads to dominance of the one-photon Fock-state $|1\rangle$ in the QD's emission with vanishing multi-photon and vacuum contributions, thus being most suitable for the on-demand single-photon generation. Simply by utilizing the Hong-Ou-Mandel (HOM) effect~\cite{Hong.Ou.ea:1987}, single and indistinguishable photons can be used to prepare a two-photon Fock-state $|2\rangle$~\cite{Nagata2007,Mueller.Vural.ea:2017,Bennett2016}.

In this work, we take this latter approach to study the slow-light effect and the acquired delay in a hot cesium (Cs) vapor~\cite{Camacho.Pack.ea:2007} for one $|1\rangle$ and two-photon Fock-states $|2\rangle$. In doing so, we extend the present studies using QDs which probed so far only single photons by delaying~\cite{Akopian.Wang.ea:2011,Vural.Portalupi.ea:2018,Trotta.Martin-Sanchez.ea:2016,Kroh2019} or polarization dependent routing~\cite{Maisch2020} in a hot vapor, to the regime of the two-photon states. 

\noindent A Cs vapor dispersive medium in combination with single QD photons recently proved highly effective for narrow bandwidth filtering by exploiting the Faraday effect~\cite{Portalupi.Widmann.ea:2016}. Moreover, strong group velocity dispersion has been key to sensitive spectroscopic characterization of QD's, particularly its spectral diffusion dynamics, by mapping frequency domain to time domain~\cite{Vural.Maisch.ea:2020}. 
As for two-photon states, the effective combination with a slow-light medium can be useful to strengthen its interferometric phase estimation with superresolution and supersensitivity~\cite{Mitchell2004,Nagata2007,Mueller.Vural.ea:2017,Bennett2016}, since the steep dispersion in a vapor additionally enhances spectral phase sensitivity~\cite{Shi.Boyd.ea:2007,Shi2008,Bortolozzo.Residori.ea:2013}.


\begin{figure*}
	\includegraphics{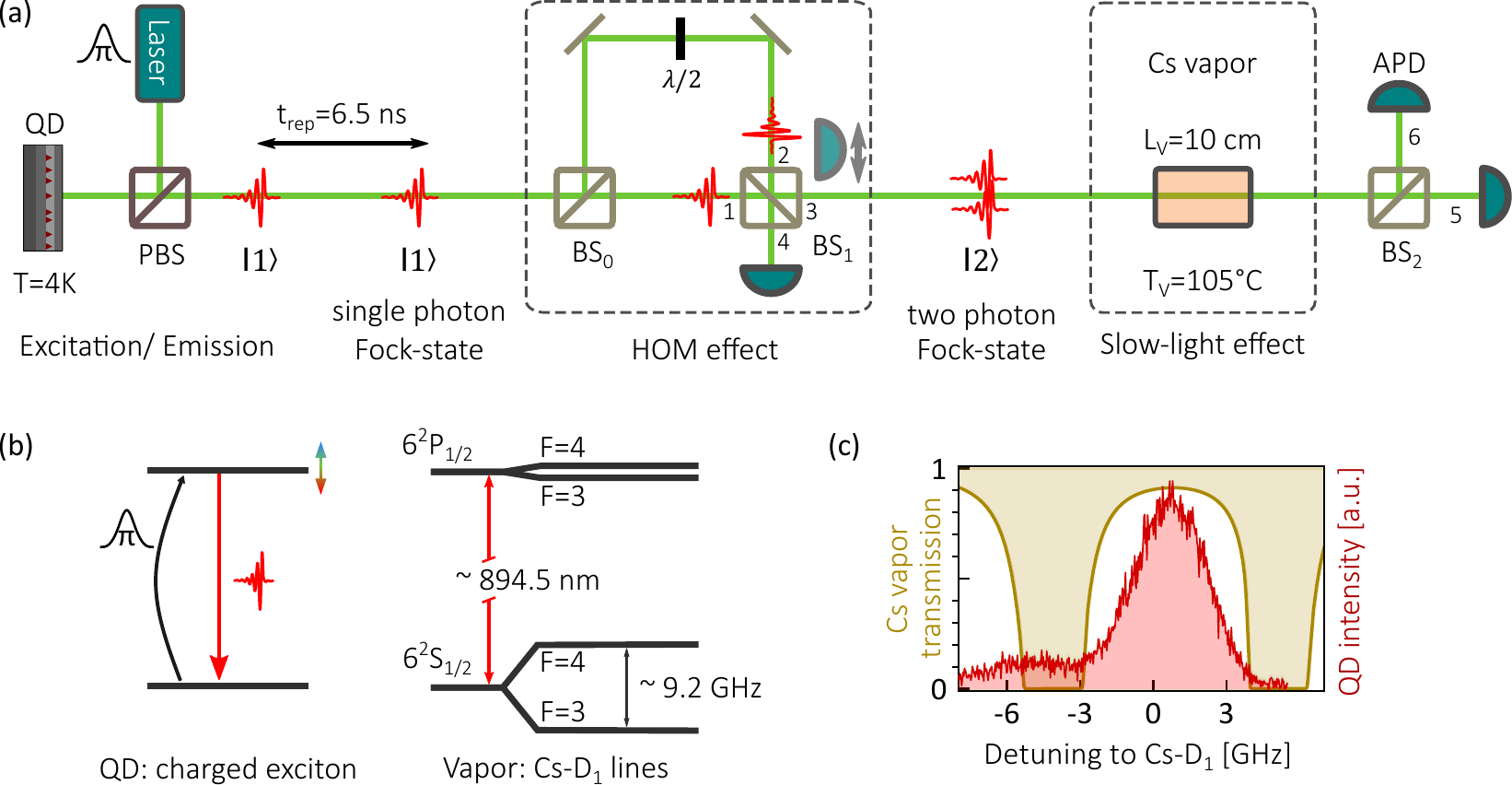}
	\caption{\textbf{Experimental scheme}. a, Sketch of the experimental setup: a quantum dot (QD) is excited by a resonant $\pi$-pulse. The emitted single photons $|1\rangle$ are fed into an unbalanced Mach-Zehnder interferometer to generate probabilisticly the two-photon Fock-state $|2\rangle$ as a result of the Hong-Ou-Mandel (HOM) effect at the beam splitter (BS$_1$). One output port (3) can be linked to a slow-light medium (here cesium (Cs) vapor) of length $L_V=10\,$cm. After the beam splitter (BS$_2$) photon-correlation and TCSPC are measured using single-photon counting modules (APD). b, Illustration of the resonant pumping of the QD's charged exciton state and the following emission of a photon at the Cs-D$_1$ transition energy. c, Simulation~\cite{Zentile.Keaveney.ea:2015} of the vapor transmission at the operated temperature of $T_V=105^\circ$C alongside with a measured emission spectrum of the QD.}
	\label{fig:setup}
\end{figure*}

\subsection*{Experimental framework} A single InGaAs QD is excited via a short optical $\pi$ pulse which is resonant to its charged exciton transition. It is specified via power and polarization dependent measurements. The wavelength of the QD transition is fine-tuned to the Cs-D$_1$ lines by the application of a uniaxial strain-field~\cite{Joens.Hafenbrak.ea:2011,Trotta.Martin-Sanchez.ea:2016} (Fig.~\ref{fig:setup}b). The resonant laser ($3\,$ps pulse length, $6.5\,$ns repetition period) is suppressed by cross-polarized excitation and detection while the QD emission is filtered to direct the zero-phonon line into an unbalanced Mach-Zehnder interferometer (Fig.~\ref{fig:setup}a). This serves to realize probabilisticly the HOM interference for successively emitted photons at the output beam splitter (BS$_1$). In fact, indistinguishable one-photon Fock-states $|1\rangle$ entering at BS$_1$ from different ports will yield $\frac{1}{\sqrt{2}}(|2,0\rangle-|0,2\rangle)$ at the outputs, that is a path-entangled N00N-state with both photons being in one port. Considering one output port allows to project out the $|2\rangle$-state and thus investigation of the two-photon Fock-state.

This state is further propagated through a hot Cs vapor to impinge on the analyzing BS$_2$. Here, a coincidence at the output ports (5 \& 6) would signify a photon pair, as opposed to the arrival of single photons. Notably, the bunching of photons due to HOM interference at BS$_1$ is here testified as doubled coincidence counts. Furthermore, successful interference is imprinted in the temporal shape of the coincidence peak~\cite{Legero.Wilk.ea:2006,Vural.Maisch.ea:2020}. By these both characteristics, the important distinction between a pair of single photons in distinguishable modes and the $|2\rangle$-state can be made~\cite{Kim2018,Rezai2018}.   

To study the vapor induced delay, time-correlated single-photon counting~(TCSPC) is performed correlating the arrival time of photons at one APD to the excitation laser pulses. As for the $|2\rangle$-state, TCSPC is heralded by a coincidence detection (at ports 5 \& 6). Only those cases are postselected which allow to infer the acquired delay for the photon pairs during the propagation through the slow-light medium. 

Fig.~\ref{fig:setup}c shows the vapor transmission spectrum at the set temperature of $T_V=105^\circ$C for a cell length of $L_V=10\,$cm which serves as the slow-light medium. At these conditions the speed of light is reduced by one order of magnitude between the Cs-D$_1$ hyperfine-split transitions. Still, the transmission window for the photons exceeds $90\,$\%.

In Fig.~\ref{fig:setup}c the vapor transmission is compared to the measured emission spectrum of the QD, which is tuned to the center of the  Cs-D$_1$ lines. The complex propagation of different spectral components in the vapor is reduced by tuning the QD to match with the center of the transmission window~\cite{Camacho.Pack.ea:2007,Vural.Portalupi.ea:2018}. The QD's spectrum is inhomogeneously broadened due to several decoherence mechanisms affecting the QD's two-level system~\cite{Kuhlmann.Houel.ea:2013,Stanley.Matthiesen.ea:2014,Schimpf.Reindl.ea:2019,Vural2020,Vural.Maisch.ea:2020}. These lead to spectral diffusion, which has been intensively studied for this particular QD in Ref.~\cite{Vural.Maisch.ea:2020}. We note that the emission of the QD in consideration has an excellent single-photon purity, confirmed by photon correlation after a beam splitter~\cite{Michler.Kiraz.ea:2000} yielding $g^{(2)}(0)=0.014\pm0.005$.  

\section{Results}
\begin{figure}
	\includegraphics[width=1.00\linewidth]{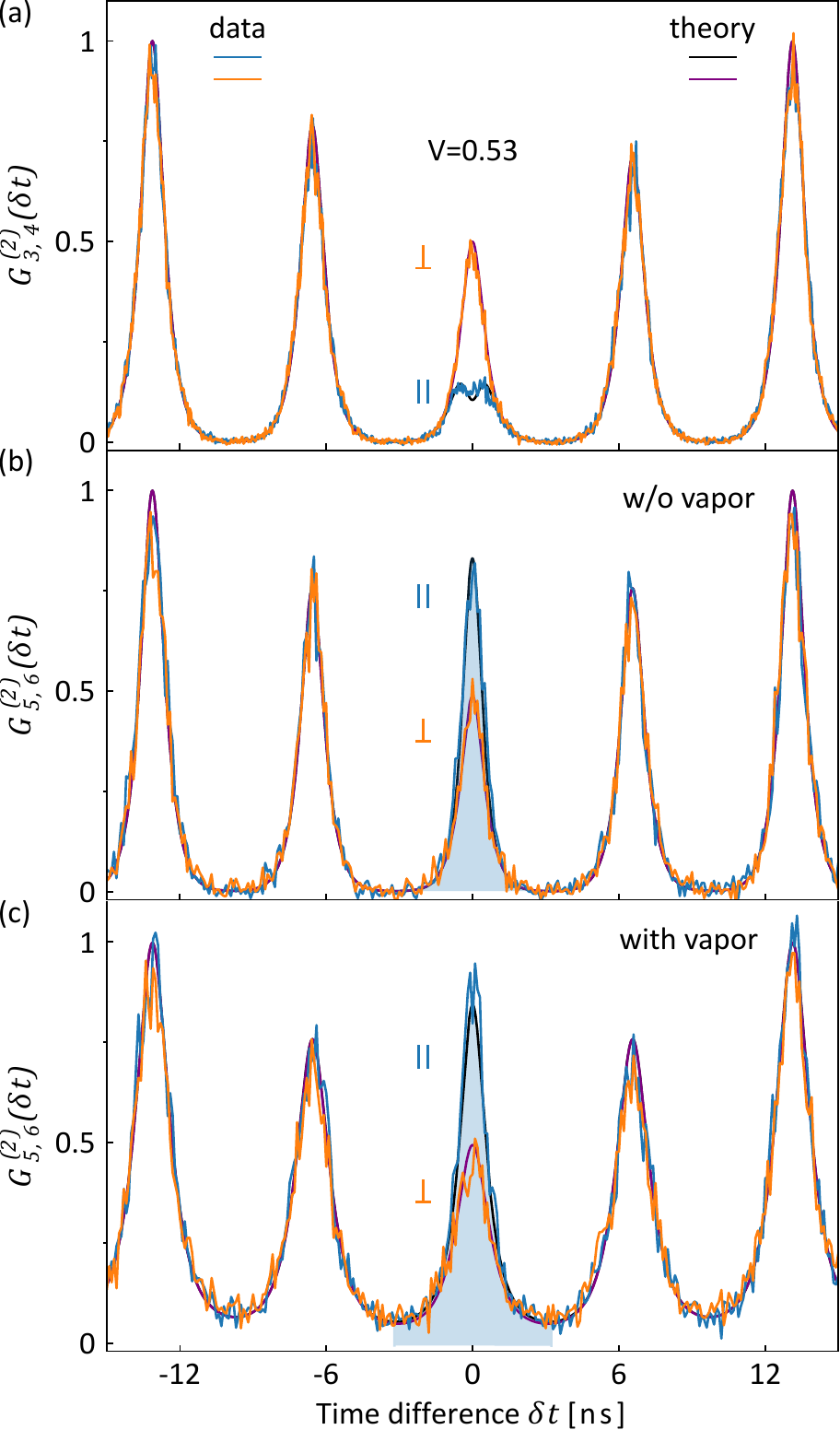}
	\caption{\textbf{Correlation measurements}. a, Hong-Ou-Mandel (HOM) interference detected at the two output ports of BS$_1$ of the HOM interferometer (see Fig.~\ref{fig:setup}a). Parallel polarized interfering photons ($\parallel$, blue) are compared to non-interfering perpendicular polarized photons ($\perp$, orange), which yields the visibility of $V=0.53$. Theory curves are plotted behind the data. b, Correlation measurement at the output ports of BS$_2$ following one HOM interferometer output port. In this constellation, interfering photons at BS$_1$ are signified by increased coincidence at BS$_2$, which yields again the same visibility. The events contained in the shadowed coincidence peak are postselected for the data in Fig.~\ref{fig:tcspc}. c, Same correlation measurement as in b, but with previous propagation of photons through the Cs vapor.
		\label{fig:hom}
	}    
\end{figure}
\subsection*{Generation of the two-photon Fock-state}

In a first step, we investigate the quantum interference at BS$_1$. Fig.~\ref{fig:hom}a shows the photon-correlation histogram for the HOM measurement when detected at ports 3 \& 4. Due to the possible paths, consecutive photons can take in the HOM setup, a peak pattern arises where only the central coincidence peak consists of the photons entering the beam splitter simultaneously from different ports (1 \& 2). When cross-polarized, non-interfering photons are investigated (orange curves), the coincidence peak amounts to half of the outermost peaks which relate to the Poissonian-level. For the parallel polarized case (blue curves), we find a strong reduction of coincidences with a narrow dip within the peak. The coalescence reveals indistinguishability of successive photons~\cite{Hong.Ou.ea:1987,Santori.Fattal.ea:2002}, while the still present modulated peak~\cite{Legero.Wilk.ea:2006,Weber.Kambs.ea:2019} is a result of broadening of the emission line. The HOM interference visibility is determined by comparing parallel and cross-polarized coincidence peak areas $V=1-G_{3,4}^{(2)}(0)_{\parallel}/G_{3,4}^{(2)}(0)_{\perp}=0.53\pm0.03$ and amounts to the percentage of successfully generated two-photon states after the beam splitter.

The observed coalescence implies both preparation of the desired 2002-state and consequently, bunching of photons in one output port. To investigate this, we link one output arm of the HOM setup to BS$_2$ for a correlation measurement at ports 5 \& 6. The projection of only one output path leaves a two-photon Fock-state $|2\rangle$ which is inspected in the central coincidence peak of the correlation histogram (Fig~\ref{fig:hom}b).

\noindent Again, starting with the cross-polarized case, we find the same pattern as at the outputs of the HOM setup. Considering the parallel case, the central coincidence peak displays a strong increase as opposed to the coincidence reduction previously observed. This demonstrates the successful propagation and detection of a $|2\rangle$-state. Indeed, the coalescence present in the HOM measurement is found to have increased the probability to find a pair of photons in one output arm (here 3). Notably, the analysis of the $|2\rangle$-state which results from the HOM interference fully reproduces the visibility. The value is again $V=|1-G_{5,6}^{(2)}(0)_{\parallel}/G_{5,6}^{(2)}(0)_{\perp}|=0.53\pm0.04$. We note that the shape of the central peak is determined by retrieving the missing coincidences of the HOM experiment as a narrow peak (see Fig.~\ref{fig:apx}, the appendix and~\cite{Rezai2018}). 

In the next step, we study the transmission and investigate whether the $|2\rangle$-state can survive the interaction with a slow-light medium. For that, the hot Cs vapor is included into the path of the photons. The correlation histogram acquired after the propagation through the vapor at ports 5 \& 6 is depicted in Fig.~\ref{fig:hom}c. It shows the same qualitative peak pattern with the same visibility as before, proving the successful propagation of the $|2\rangle$-state through the slow-light medium. Indeed, this is in accordance with a previous study~\cite{Vural.Portalupi.ea:2018}, where the interaction of single photons with a linear dispersive medium was shown to preserve quantum interference. However, the peaks in the histogram display broadening as a result of pulse distortion that is closely connected to the vapor response and the presence of QD spectral diffusion~\cite{Vural.Maisch.ea:2020}. It is worth to mention, that the effect of detector jitter is reduced due to this temporal broadening. In particular, the temporal shape of the central coincidence peak which signifies the $|2\rangle$-state remains obviously narrower than all other peaks of correlations from temporally separate single photons. The apparent implication is that the coincidence peak reaches closer to the theoretical maximum (see Fig.~\ref{fig:apx}).

\begin{figure}
	\includegraphics[width=1.00\linewidth]{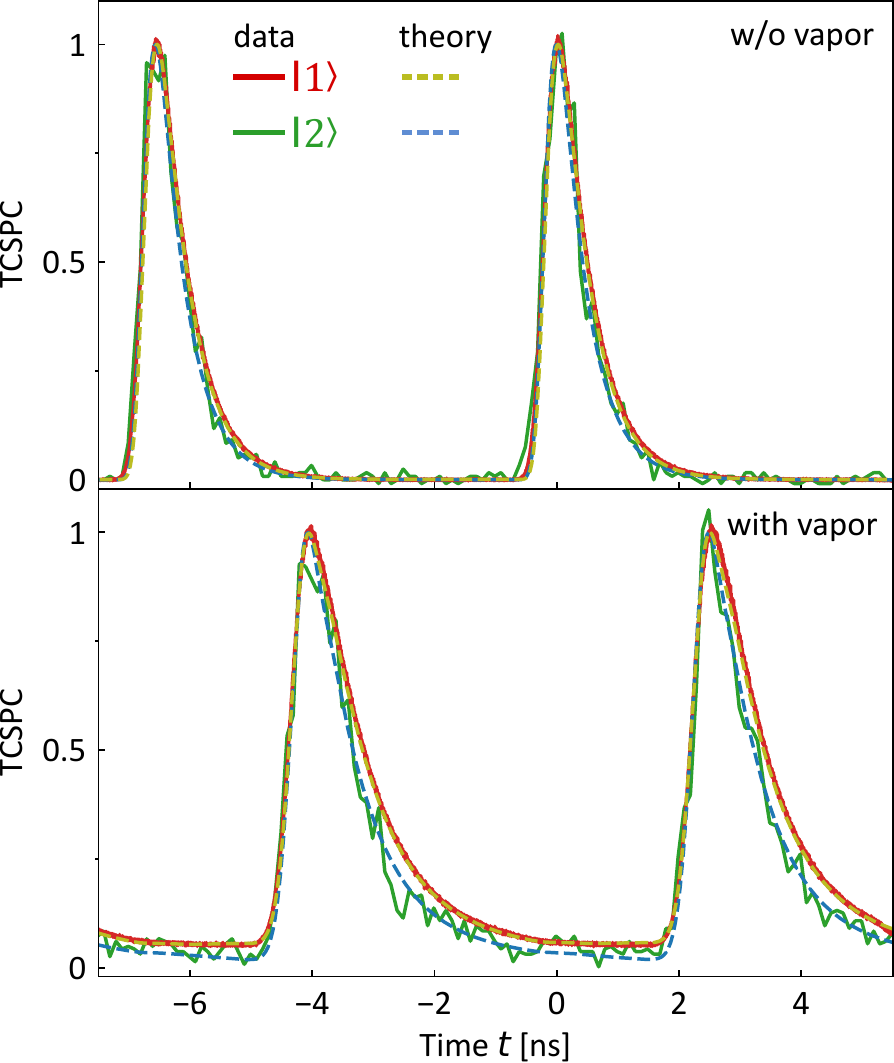}
	\caption{\textbf{Delayed Fock-states}. a,  Time-correlated single-photon counting (TCSPC) for the one-photon Fock-state $|1\rangle$ (red curve, $1\,$ps binning, all detection events) and the two-photon Fock-state $|2\rangle$ (green line, $100\,$ps binning, postselected on events within the highlighted coincidence peak in Fig.~\ref{fig:hom}b \& c) detected at port 6, without vapor after the HOM setup. Dashed yellow curve is an exponential fit (decay constant $\tau=0.43\,$ns) under consideration of the APD response. Dashed blue curve is a simulation for two-photon states. b, TCSPC with the propagation of photons through the Cs vapor. Color code as in a.}
	\label{fig:tcspc}
\end{figure}

\subsection*{Temporal delay for Fock-states} Having shown the successful generation of photon pairs in the $|2\rangle$-state outlasting the interaction with the vapor, we now investigate the delay by means of TCSPC for the two Fock-states $|1\rangle$ and $|2\rangle$ under consideration.

\noindent Fig.~\ref{fig:tcspc}a shows TCSPC detected at port 6 of BS$_2$ for the case of parallel polarized photons. Postselection on a coincidence event within the highlighted bunching peaks of Fig.~\ref{fig:hom}b delivers TCSPC for the $|2\rangle$-state. Taking all other detection events yield TCSPC for the $|1\rangle$-state. We find the detection times and the temporal shapes of both photon states to be similar. As for the $|2\rangle$-state, a slightly modulated temporal shape is anticipated due to two-photon interference, which induces beating~\cite{Legero.Wilk.ea:2006}. However, the severe limitation of detection rates due to the postselection results in a higher statistical deviation in each bin preventing a clear distinction of the temporal shapes of the states.  

\noindent Fig.~\ref{fig:tcspc}b shows TCSPC measurements after the propagation of parallel polarized photons through the Cs vapor using the same detection procedure as before. The arrival time of photons is now delayed by $\sim 3\,$ns due to decreased group velocity in the atomic vapor for the photons of both states. Moreover, they display a slight pulse distortion due to chromatic dispersion of the vapor~\cite{Vural.Portalupi.ea:2018}. Once again higher statistical deviation in each bin is present for the data of postselected events of the $|2\rangle$-state. However, the temporal broadening due to the vapor is beneficial to distinguish the temporal shapes of the states. The two-photon interference yields a narrower temporal form in TCSPC as it does in the correlation measurements under emission line broadening. Simulations (dashed curves) that take the known spectral diffusion process for this QD~\cite{Vural.Maisch.ea:2020} into account, reproduce the data faithfully. It is important to note, that for a Fourier-limited emitter, the temporal shapes would not differ. In this respect, the correlation measurements at a beam splitter are a mandatory tool to distinguish between the Fock-states. 

\section{Discussion}
In summary, we utilized the bunching of indistinguishable photons at a beam splitter, as a result of the HOM interference, to generate a two-photon Fock-state. We verified the accordance of coalescence in the HOM measurement with the bunching of photons in one of the output ports by a correlation measurement on the two-photon Fock-state. The interaction of the two-photon Fock-state with a dispersive cesium vapor compares to the single photon case as signified by the same acquired delay and temporal form in TCSPC measurement. Thus, we experimentally verified that higher number Fock-state ($N=2$) shows no hindrance to be utilized in slow-light media. This combination can open new perspectives in interferometry, reinforcing the enhancement in spectral phase sensitivity provided by the slow-light effect~\cite{Shi.Boyd.ea:2007,Shi2008,Bortolozzo.Residori.ea:2013} with the enabling feature of the two-photon Fock-state for phase measurements with superresolution and supersensitivity~\cite{Mitchell2004,Nagata2007,Mueller.Vural.ea:2017,Bennett2016}. 

\begin{acknowledgments}
	We acknowledge financial support from the DFG (MI 500/31-1). S. L. P. greatly acknowledges the BW Stiftung \lq\lq Post-doc Eliteprogramm\rq\rq \ via the project Hybrideye.
	We thank Robert L\"ow, Julian Maisch and Jonas Weber for fruitful discussions.
\end{acknowledgments}
\section*{appendix}
\begin{figure}
	\includegraphics[width=1.00\linewidth]{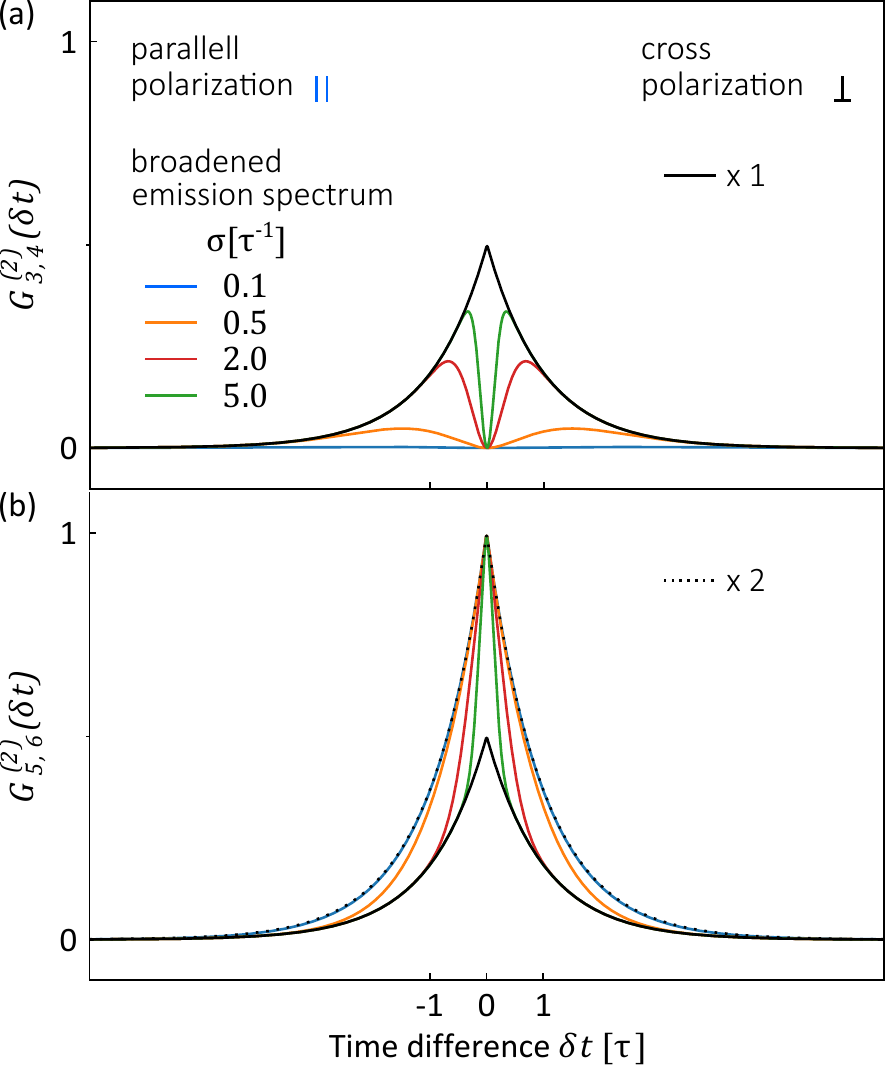}
	\caption{\textbf{Theory:} a, HOM interference at BS$_1$ in the presence of line broadening of the emission with standard deviation $\sigma$ of broadened spectrum as specified in Eq.~\eqref{eq:HOM}. The black curve represents the non-interfering case of cross polarized photons. b, Photon correlation at BS$_2$ for the same cases as considered in a. Dashed black curve shows twice the cross polarized curve for comparison. The coincidences reduced in a are cumulated in the coincidence of b. In the actual experiment, detector jitter prohibits the narrow dip from reaching zero and the narrow peak from one.}
	\label{fig:apx}
\end{figure}
A single photon Fock-state with wavepacket $\chi$ can be described~\cite{Loudon2000} as $|1\rangle=\int d\omega\, \chi(\omega) \hat{a}^\dagger(\omega) |0\rangle$. Here $\hat{a}^\dagger(\omega)$ is a bosonic creation operator at a single angular frequency $\omega$. Using the Fourier-transformation $\mathcal{F}\mathcal{T}$ yields the temporal wavepacket:
\begin{equation}
\chi(t)=\mathcal{F}\mathcal{T}\left\{\chi(\omega)\right\}=\int \frac{d\omega}{\sqrt{2\pi}} e^{-i\omega t}\chi(\omega) \quad.
\end{equation}

The propagation of a photon pulse in a dispersive medium leads to a change in its phase via the transformation:
\begin{equation}
\hat{a}(\omega) \mapsto  e^{-i \omega \frac{L n(\omega)}{c}} \hat{a}(\omega) \quad,
\label{eq:vapor}
\end{equation}  
where $L$ is the propagation distance, $c$ the speed of light and $n(\omega)$ is the index of refraction. Please note, that in the detection process ${n(\omega)\mapsto n(\omega)+\frac{i}{2 \omega/c} \alpha(\omega)}$ will be used, where the absorption according to the Beer-Lambert's law, induced by the coefficient $\alpha(\omega)$, appears as a frequency dependent detection efficiency, hence does not hurt the quantum mechanical description. 

The general amplitude spectrum of a Lorentzian photon generated as a result of spontaneous emission after propagation along $L$ in the vapor (indicated by the superscript) takes the form:
\begin{equation}
\label{eq:ampspec}
\chi^L_0(\omega)=\sqrt{\frac{2\tau}{\pi}}\frac{e^{i\omega t_0}}{1-2i\tau(\omega-\omega_0)}e^{i \omega \frac{L n(\omega)}{c}} \quad,
\end{equation}
where $t_0$ is the time of photon generation, $\tau$ the decay constant and $\omega_0$ the carrier frequency of the wavepacket which will be indicated by the subscript.

 The Hong-Ou-Mandel interference~\cite{Hong.Ou.ea:1987,Santori.Fattal.ea:2002} can be studied by photon-correlation measurements at different outputs $G_{3,4}^{(2)}(\delta t)$ or at one output $G_{3,3}^{(2)}(\delta t)$ in a time-resolved manner. Following the description in~\cite{Legero.Wilk.ea:2006} the state of the single photons $|\Psi\rangle=|1\chi_1^0\rangle_1\otimes|1\chi_2^0\rangle_2$ entering BS$_1$ from the spatial modes indicated by the subscript of the kets yield: 
\begin{equation}
G_{3,4}^{(2)}(\delta t)
\propto\int dt \frac{1}{4}|\chi_1^{0}(t+\delta t)\chi_2^{0}(t)-\chi_1^{0}(t)\chi_2^{0}(t+\delta t)|^2
\end{equation}
\begin{equation}
G_{3,3}^{(2)}(\delta t)
\propto\int dt \frac{1}{4}|\chi_1^{0}(t+\delta t)\chi_2^{0}(t)+\chi_1^{0}(t)\chi_2^{0}(t+\delta t)|^2
\end{equation}
where $G_{i,j}^{(2)}(\delta t)=\int dt \langle\Psi|\hat{a}_i(t)\hat{a}_j(t+\delta t)\hat{a}^\dagger_j(t+\delta t)\hat{a}^\dagger _i(t)|\Psi\rangle$.

If the photons posses orthogonal polarization $H$ and $V$:
\begin{equation}
G_{i,j}^{(2)}(\delta t)
\propto\int dt \frac{1}{4}|\chi_1^{0}(t+\delta t)\chi_2^{0}(t)|^2+\frac{1}{4}|\chi_1^{0}(t)\chi_2^{0}(t+\delta t)|^2 
\end{equation}
thus the correlation displays intensities only and no interference occurs as expected.

For the ensemble average of the broadened emission spectrum~\cite{Vural.Maisch.ea:2020} in case of a simple Normal distribution $\mathcal{N}$ with variance $\sigma^2$, the integrated histograms are obtained as:
\begin{align}
\langle G_{i,j}^{(2)}(\delta t)\rangle_{\mathcal{N}(0,2\sigma^2)}=\frac{1}{4\tau}e^{-|\delta t|/\tau}\left(1\mp e^{-\sigma^2\delta t^2}\right) \quad,
\label{eq:HOM}
\end{align}
where $\langle G_{i,j}^{(2)}(\delta t)\rangle_f=\int d(\delta\omega) f(\delta\omega) G_{i,j}^{(2)}(\delta t)$, $\delta \omega$ integrated over the carrier frequency differences of the photons, while the minus (plus) in the brackets is to be taken for the case of different (same) output modes. Despite broadening of the emission spectrum, one can notice that coincidence at different outputs still completely vanishes for $\delta t=0$, and hence bunching in one output mode is present. Fig.~\ref{fig:apx} illustrates this for various Gaussian broadened emission, revealing the narrow dip or peak, respectively. 

In our experiments, we used an additional beam splitter to measure the correlation which is present in one output arm of the HOM setup (port 3). This allows to acquire the same histogram with coincidence rate being the only compromise, since the correlation at ports (5 \& 6) at the BS$_2$ produce the same histogram i.e. $G_{3,3}^{(2)}(\delta t)\propto G_{5,6}^{(2)}(\delta t)$. For the vapor in the path of the photons, relations \eqref{eq:vapor} and \eqref{eq:ampspec} were used. The theory curves to fit the data used spectral diffusion parameters of a previous study on the same QD~\cite{Vural.Maisch.ea:2020}.

TCSPC is calculated via $\langle\Psi|\hat{a}_6(t)\hat{a}^\dagger_6(t)|\Psi\rangle$ with  $|\Psi\rangle=|1\chi_1^L\rangle_6$ and ${|\Psi\rangle=\frac{1}{\sqrt{2}}(|1\chi_1^L\rangle_5\otimes|1\chi_2^L\rangle_6+|1\chi_2^L\rangle_5\otimes|1\chi_1^L\rangle_6)}$ for one and two-photon states postselected on a coincidence, respectively. Additionally, the spectral diffusion process is accounted for.


\bibliography{references_fock}

\begin{thebibliography}{42}%
\makeatletter
\providecommand \@ifxundefined [1]{%
 \@ifx{#1\undefined}
}%
\providecommand \@ifnum [1]{%
 \ifnum #1\expandafter \@firstoftwo
 \else \expandafter \@secondoftwo
 \fi
}%
\providecommand \@ifx [1]{%
 \ifx #1\expandafter \@firstoftwo
 \else \expandafter \@secondoftwo
 \fi
}%
\providecommand \natexlab [1]{#1}%
\providecommand \enquote  [1]{``#1''}%
\providecommand \bibnamefont  [1]{#1}%
\providecommand \bibfnamefont [1]{#1}%
\providecommand \citenamefont [1]{#1}%
\providecommand \href@noop [0]{\@secondoftwo}%
\providecommand \href [0]{\begingroup \@sanitize@url \@href}%
\providecommand \@href[1]{\@@startlink{#1}\@@href}%
\providecommand \@@href[1]{\endgroup#1\@@endlink}%
\providecommand \@sanitize@url [0]{\catcode `\\12\catcode `\$12\catcode
  `\&12\catcode `\#12\catcode `\^12\catcode `\_12\catcode `\%12\relax}%
\providecommand \@@startlink[1]{}%
\providecommand \@@endlink[0]{}%
\providecommand \url  [0]{\begingroup\@sanitize@url \@url }%
\providecommand \@url [1]{\endgroup\@href {#1}{\urlprefix }}%
\providecommand \urlprefix  [0]{URL }%
\providecommand \Eprint [0]{\href }%
\providecommand \doibase [0]{https://doi.org/}%
\providecommand \selectlanguage [0]{\@gobble}%
\providecommand \bibinfo  [0]{\@secondoftwo}%
\providecommand \bibfield  [0]{\@secondoftwo}%
\providecommand \translation [1]{[#1]}%
\providecommand \BibitemOpen [0]{}%
\providecommand \bibitemStop [0]{}%
\providecommand \bibitemNoStop [0]{.\EOS\space}%
\providecommand \EOS [0]{\spacefactor3000\relax}%
\providecommand \BibitemShut  [1]{\csname bibitem#1\endcsname}%
\let\auto@bib@innerbib\@empty
\bibitem [{\citenamefont {Michler}(2017)}]{Michler:2017}%
  \BibitemOpen
  \bibinfo {editor} {\bibfnamefont {P.}~\bibnamefont {Michler}},\ ed.,\
  \href@noop {} {\emph {\bibinfo {title} {{Q}uantum Dots for {Q}uantum
  {I}nformation {T}echnologies}}}\ (\bibinfo  {publisher} {Springer
  International Publishing},\ \bibinfo {year} {2017})\BibitemShut {NoStop}%
\bibitem [{\citenamefont {He}\ \emph {et~al.}(2013)\citenamefont {He},
  \citenamefont {He}, \citenamefont {Wei}, \citenamefont {Wu}, \citenamefont
  {Atat\"ure}, \citenamefont {Schneider}, \citenamefont {H\"ofling},
  \citenamefont {Kamp}, \citenamefont {Lu},\ and\ \citenamefont
  {Pan}}]{He.He.ea:2013}%
  \BibitemOpen
  \bibfield  {author} {\bibinfo {author} {\bibfnamefont {Y.-M.}\ \bibnamefont
  {He}}, \bibinfo {author} {\bibfnamefont {Y.}~\bibnamefont {He}}, \bibinfo
  {author} {\bibfnamefont {Y.-J.}\ \bibnamefont {Wei}}, \bibinfo {author}
  {\bibfnamefont {D.}~\bibnamefont {Wu}}, \bibinfo {author} {\bibfnamefont
  {M.}~\bibnamefont {Atat\"ure}}, \bibinfo {author} {\bibfnamefont
  {C.}~\bibnamefont {Schneider}}, \bibinfo {author} {\bibfnamefont
  {S.}~\bibnamefont {H\"ofling}}, \bibinfo {author} {\bibfnamefont
  {M.}~\bibnamefont {Kamp}}, \bibinfo {author} {\bibfnamefont {C.-Y.}\
  \bibnamefont {Lu}},\ and\ \bibinfo {author} {\bibfnamefont {J.-W.}\
  \bibnamefont {Pan}},\ }\href {http://dx.doi.org/10.1038/nnano.2012.262}
  {\bibfield  {journal} {\bibinfo  {journal} {Nature Nano.}\ }\textbf {\bibinfo
  {volume} {8}},\ \bibinfo {pages} {213} (\bibinfo {year} {2013})}\BibitemShut
  {NoStop}%
\bibitem [{\citenamefont {M\"uller}\ \emph {et~al.}(2014)\citenamefont
  {M\"uller}, \citenamefont {Bounouar}, \citenamefont {J\"ons}, \citenamefont
  {Gl\"assl},\ and\ \citenamefont {Michler}}]{Muller.Bounouar.ea:2014}%
  \BibitemOpen
  \bibfield  {author} {\bibinfo {author} {\bibfnamefont {M.}~\bibnamefont
  {M\"uller}}, \bibinfo {author} {\bibfnamefont {S.}~\bibnamefont {Bounouar}},
  \bibinfo {author} {\bibfnamefont {K.~D.}\ \bibnamefont {J\"ons}}, \bibinfo
  {author} {\bibfnamefont {M.}~\bibnamefont {Gl\"assl}},\ and\ \bibinfo
  {author} {\bibfnamefont {P.}~\bibnamefont {Michler}},\ }\href
  {http://dx.doi.org/10.1038/nphoton.2013.377} {\bibfield  {journal} {\bibinfo
  {journal} {Nature Photon.}\ }\textbf {\bibinfo {volume} {8}},\ \bibinfo
  {pages} {224} (\bibinfo {year} {2014})}\BibitemShut {NoStop}%
\bibitem [{\citenamefont {Liu}\ \emph {et~al.}(2019)\citenamefont {Liu},
  \citenamefont {Su}, \citenamefont {Wei}, \citenamefont {Yao}, \citenamefont
  {Silva}, \citenamefont {Yu}, \citenamefont {Iles-Smith}, \citenamefont
  {Srinivasan}, \citenamefont {Rastelli}, \citenamefont {Li},\ and\
  \citenamefont {Wang}}]{Liu.Su.ea:2019}%
  \BibitemOpen
  \bibfield  {author} {\bibinfo {author} {\bibfnamefont {J.}~\bibnamefont
  {Liu}}, \bibinfo {author} {\bibfnamefont {R.}~\bibnamefont {Su}}, \bibinfo
  {author} {\bibfnamefont {Y.}~\bibnamefont {Wei}}, \bibinfo {author}
  {\bibfnamefont {B.}~\bibnamefont {Yao}}, \bibinfo {author} {\bibfnamefont
  {S.~F. C.~d.}\ \bibnamefont {Silva}}, \bibinfo {author} {\bibfnamefont
  {Y.}~\bibnamefont {Yu}}, \bibinfo {author} {\bibfnamefont {J.}~\bibnamefont
  {Iles-Smith}}, \bibinfo {author} {\bibfnamefont {K.}~\bibnamefont
  {Srinivasan}}, \bibinfo {author} {\bibfnamefont {A.}~\bibnamefont
  {Rastelli}}, \bibinfo {author} {\bibfnamefont {J.}~\bibnamefont {Li}},\ and\
  \bibinfo {author} {\bibfnamefont {X.}~\bibnamefont {Wang}},\ }\href
  {https://doi.org/10.1038/s41565-019-0435-9} {\bibfield  {journal} {\bibinfo
  {journal} {Nature Nanotechnology}\ ,\ } (\bibinfo {year} {2019})}\BibitemShut
  {NoStop}%
\bibitem [{\citenamefont {Wang}\ \emph {et~al.}(2019)\citenamefont {Wang},
  \citenamefont {Qin}, \citenamefont {Ding}, \citenamefont {Chen},
  \citenamefont {Chen}, \citenamefont {You}, \citenamefont {He}, \citenamefont
  {Jiang}, \citenamefont {You}, \citenamefont {Wang}, \citenamefont
  {Schneider}, \citenamefont {Renema}, \citenamefont {H\"ofling}, \citenamefont
  {Lu},\ and\ \citenamefont {Pan}}]{Wang.Qin.ea:2019}%
  \BibitemOpen
  \bibfield  {author} {\bibinfo {author} {\bibfnamefont {H.}~\bibnamefont
  {Wang}}, \bibinfo {author} {\bibfnamefont {J.}~\bibnamefont {Qin}}, \bibinfo
  {author} {\bibfnamefont {X.}~\bibnamefont {Ding}}, \bibinfo {author}
  {\bibfnamefont {M.-C.}\ \bibnamefont {Chen}}, \bibinfo {author}
  {\bibfnamefont {S.}~\bibnamefont {Chen}}, \bibinfo {author} {\bibfnamefont
  {X.}~\bibnamefont {You}}, \bibinfo {author} {\bibfnamefont {Y.-M.}\
  \bibnamefont {He}}, \bibinfo {author} {\bibfnamefont {X.}~\bibnamefont
  {Jiang}}, \bibinfo {author} {\bibfnamefont {L.}~\bibnamefont {You}}, \bibinfo
  {author} {\bibfnamefont {Z.}~\bibnamefont {Wang}}, \bibinfo {author}
  {\bibfnamefont {C.}~\bibnamefont {Schneider}}, \bibinfo {author}
  {\bibfnamefont {J.~J.}\ \bibnamefont {Renema}}, \bibinfo {author}
  {\bibfnamefont {S.}~\bibnamefont {H\"ofling}}, \bibinfo {author}
  {\bibfnamefont {C.-Y.}\ \bibnamefont {Lu}},\ and\ \bibinfo {author}
  {\bibfnamefont {J.-W.}\ \bibnamefont {Pan}},\ }\href
  {https://doi.org/10.1103/PhysRevLett.123.250503} {\bibfield  {journal}
  {\bibinfo  {journal} {Phys. Rev. Lett.}\ }\textbf {\bibinfo {volume} {123}},\
  \bibinfo {pages} {250503} (\bibinfo {year} {2019})}\BibitemShut {NoStop}%
\bibitem [{\citenamefont {Nagata}\ \emph {et~al.}(2007)\citenamefont {Nagata},
  \citenamefont {Okamoto}, \citenamefont {O{\textquoteright}Brien},
  \citenamefont {Sasaki},\ and\ \citenamefont {Takeuchi}}]{Nagata2007}%
  \BibitemOpen
  \bibfield  {author} {\bibinfo {author} {\bibfnamefont {T.}~\bibnamefont
  {Nagata}}, \bibinfo {author} {\bibfnamefont {R.}~\bibnamefont {Okamoto}},
  \bibinfo {author} {\bibfnamefont {J.~L.}\ \bibnamefont
  {O{\textquoteright}Brien}}, \bibinfo {author} {\bibfnamefont
  {K.}~\bibnamefont {Sasaki}},\ and\ \bibinfo {author} {\bibfnamefont
  {S.}~\bibnamefont {Takeuchi}},\ }\href
  {https://doi.org/10.1126/science.1138007} {\bibfield  {journal} {\bibinfo
  {journal} {Science}\ }\textbf {\bibinfo {volume} {316}},\ \bibinfo {pages}
  {726} (\bibinfo {year} {2007})}\BibitemShut {NoStop}%
\bibitem [{\citenamefont {Mitchell}\ \emph {et~al.}(2004)\citenamefont
  {Mitchell}, \citenamefont {Lundeen},\ and\ \citenamefont
  {Steinberg}}]{Mitchell2004}%
  \BibitemOpen
  \bibfield  {author} {\bibinfo {author} {\bibfnamefont {M.~W.}\ \bibnamefont
  {Mitchell}}, \bibinfo {author} {\bibfnamefont {J.~S.}\ \bibnamefont
  {Lundeen}},\ and\ \bibinfo {author} {\bibfnamefont {A.~M.}\ \bibnamefont
  {Steinberg}},\ }\href {https://doi.org/10.1038/nature02493} {\bibfield
  {journal} {\bibinfo  {journal} {Nature}\ }\textbf {\bibinfo {volume} {429}},\
  \bibinfo {pages} {161} (\bibinfo {year} {2004})}\BibitemShut {NoStop}%
\bibitem [{\citenamefont {M\"uller}\ \emph {et~al.}(2017)\citenamefont
  {M\"uller}, \citenamefont {Vural}, \citenamefont {Schneider}, \citenamefont
  {Rastelli}, \citenamefont {Schmidt}, \citenamefont {H\"ofling},\ and\
  \citenamefont {Michler}}]{Mueller.Vural.ea:2017}%
  \BibitemOpen
  \bibfield  {author} {\bibinfo {author} {\bibfnamefont {M.}~\bibnamefont
  {M\"uller}}, \bibinfo {author} {\bibfnamefont {H.}~\bibnamefont {Vural}},
  \bibinfo {author} {\bibfnamefont {C.}~\bibnamefont {Schneider}}, \bibinfo
  {author} {\bibfnamefont {A.}~\bibnamefont {Rastelli}}, \bibinfo {author}
  {\bibfnamefont {O.~G.}\ \bibnamefont {Schmidt}}, \bibinfo {author}
  {\bibfnamefont {S.}~\bibnamefont {H\"ofling}},\ and\ \bibinfo {author}
  {\bibfnamefont {P.}~\bibnamefont {Michler}},\ }\href
  {https://doi.org/10.1103/PhysRevLett.118.257402} {\bibfield  {journal}
  {\bibinfo  {journal} {Phys. Rev. Lett.}\ }\textbf {\bibinfo {volume} {118}},\
  \bibinfo {pages} {257402} (\bibinfo {year} {2017})}\BibitemShut {NoStop}%
\bibitem [{\citenamefont {Jennewein}\ \emph {et~al.}(2000)\citenamefont
  {Jennewein}, \citenamefont {Simon}, \citenamefont {Weihs}, \citenamefont
  {Weinfurter},\ and\ \citenamefont {Zeilinger}}]{Jennewein2000}%
  \BibitemOpen
  \bibfield  {author} {\bibinfo {author} {\bibfnamefont {T.}~\bibnamefont
  {Jennewein}}, \bibinfo {author} {\bibfnamefont {C.}~\bibnamefont {Simon}},
  \bibinfo {author} {\bibfnamefont {G.}~\bibnamefont {Weihs}}, \bibinfo
  {author} {\bibfnamefont {H.}~\bibnamefont {Weinfurter}},\ and\ \bibinfo
  {author} {\bibfnamefont {A.}~\bibnamefont {Zeilinger}},\ }\href
  {https://doi.org/10.1103/PhysRevLett.84.4729} {\bibfield  {journal} {\bibinfo
   {journal} {Phys. Rev. Lett.}\ }\textbf {\bibinfo {volume} {84}},\ \bibinfo
  {pages} {4729} (\bibinfo {year} {2000})}\BibitemShut {NoStop}%
\bibitem [{\citenamefont {Simon}\ \emph {et~al.}(2007)\citenamefont {Simon},
  \citenamefont {de~Riedmatten}, \citenamefont {Afzelius}, \citenamefont
  {Sangouard}, \citenamefont {Zbinden},\ and\ \citenamefont
  {Gisin}}]{Simon2007}%
  \BibitemOpen
  \bibfield  {author} {\bibinfo {author} {\bibfnamefont {C.}~\bibnamefont
  {Simon}}, \bibinfo {author} {\bibfnamefont {H.}~\bibnamefont
  {de~Riedmatten}}, \bibinfo {author} {\bibfnamefont {M.}~\bibnamefont
  {Afzelius}}, \bibinfo {author} {\bibfnamefont {N.}~\bibnamefont {Sangouard}},
  \bibinfo {author} {\bibfnamefont {H.}~\bibnamefont {Zbinden}},\ and\ \bibinfo
  {author} {\bibfnamefont {N.}~\bibnamefont {Gisin}},\ }\href
  {https://doi.org/10.1103/PhysRevLett.98.190503} {\bibfield  {journal}
  {\bibinfo  {journal} {Phys. Rev. Lett.}\ }\textbf {\bibinfo {volume} {98}},\
  \bibinfo {pages} {190503} (\bibinfo {year} {2007})}\BibitemShut {NoStop}%
\bibitem [{\citenamefont {Upton}\ \emph {et~al.}(2013)\citenamefont {Upton},
  \citenamefont {Harpham}, \citenamefont {Suzer}, \citenamefont {Richter},
  \citenamefont {Mukamel},\ and\ \citenamefont {Goodson}}]{Upton2013}%
  \BibitemOpen
  \bibfield  {author} {\bibinfo {author} {\bibfnamefont {L.}~\bibnamefont
  {Upton}}, \bibinfo {author} {\bibfnamefont {M.}~\bibnamefont {Harpham}},
  \bibinfo {author} {\bibfnamefont {O.}~\bibnamefont {Suzer}}, \bibinfo
  {author} {\bibfnamefont {M.}~\bibnamefont {Richter}}, \bibinfo {author}
  {\bibfnamefont {S.}~\bibnamefont {Mukamel}},\ and\ \bibinfo {author}
  {\bibfnamefont {T.}~\bibnamefont {Goodson}},\ }\href
  {https://doi.org/10.1021/jz400851d} {\bibfield  {journal} {\bibinfo
  {journal} {The Journal of Physical Chemistry Letters}\ }\textbf {\bibinfo
  {volume} {4}},\ \bibinfo {pages} {2046} (\bibinfo {year} {2013})}\BibitemShut
  {NoStop}%
\bibitem [{\citenamefont {del Valle}\ \emph {et~al.}(2011)\citenamefont {del
  Valle}, \citenamefont {Gonzalez{\textendash}Tudela}, \citenamefont
  {Cancellieri}, \citenamefont {Laussy},\ and\ \citenamefont
  {Tejedor}}]{Valle2011}%
  \BibitemOpen
  \bibfield  {author} {\bibinfo {author} {\bibfnamefont {E.}~\bibnamefont {del
  Valle}}, \bibinfo {author} {\bibfnamefont {A.}~\bibnamefont
  {Gonzalez{\textendash}Tudela}}, \bibinfo {author} {\bibfnamefont
  {E.}~\bibnamefont {Cancellieri}}, \bibinfo {author} {\bibfnamefont {F.~P.}\
  \bibnamefont {Laussy}},\ and\ \bibinfo {author} {\bibfnamefont
  {C.}~\bibnamefont {Tejedor}},\ }\href
  {https://doi.org/10.1088/1367-2630/13/11/113014} {\bibfield  {journal}
  {\bibinfo  {journal} {New Journal of Physics}\ }\textbf {\bibinfo {volume}
  {13}},\ \bibinfo {pages} {113014} (\bibinfo {year} {2011})}\BibitemShut
  {NoStop}%
\bibitem [{\citenamefont {Mu\~{n}oz}\ \emph {et~al.}(2014)\citenamefont
  {Mu\~{n}oz}, \citenamefont {del Valle}, \citenamefont {Tudela}, \citenamefont
  {M\"uller}, \citenamefont {Lichtmannecker}, \citenamefont {Kaniber},
  \citenamefont {Tejedor}, \citenamefont {Finley},\ and\ \citenamefont
  {Laussy}}]{Munoz2014}%
  \BibitemOpen
  \bibfield  {author} {\bibinfo {author} {\bibfnamefont {C.~S.}\ \bibnamefont
  {Mu\~{n}oz}}, \bibinfo {author} {\bibfnamefont {E.}~\bibnamefont {del
  Valle}}, \bibinfo {author} {\bibfnamefont {A.~G.}\ \bibnamefont {Tudela}},
  \bibinfo {author} {\bibfnamefont {K.}~\bibnamefont {M\"uller}}, \bibinfo
  {author} {\bibfnamefont {S.}~\bibnamefont {Lichtmannecker}}, \bibinfo
  {author} {\bibfnamefont {M.}~\bibnamefont {Kaniber}}, \bibinfo {author}
  {\bibfnamefont {C.}~\bibnamefont {Tejedor}}, \bibinfo {author} {\bibfnamefont
  {J.~J.}\ \bibnamefont {Finley}},\ and\ \bibinfo {author} {\bibfnamefont
  {F.~P.}\ \bibnamefont {Laussy}},\ }\href
  {https://doi.org/10.1038/nphoton.2014.114} {\bibfield  {journal} {\bibinfo
  {journal} {Nature Photonics}\ }\textbf {\bibinfo {volume} {8}},\ \bibinfo
  {pages} {550} (\bibinfo {year} {2014})}\BibitemShut {NoStop}%
\bibitem [{\citenamefont {Schumacher}\ \emph {et~al.}(2012)\citenamefont
  {Schumacher}, \citenamefont {F\"{o}rstner}, \citenamefont {Zrenner},
  \citenamefont {Florian}, \citenamefont {Gies}, \citenamefont {Gartner},\ and\
  \citenamefont {Jahnke}}]{Schumacher2012}%
  \BibitemOpen
  \bibfield  {author} {\bibinfo {author} {\bibfnamefont {S.}~\bibnamefont
  {Schumacher}}, \bibinfo {author} {\bibfnamefont {J.}~\bibnamefont
  {F\"{o}rstner}}, \bibinfo {author} {\bibfnamefont {A.}~\bibnamefont
  {Zrenner}}, \bibinfo {author} {\bibfnamefont {M.}~\bibnamefont {Florian}},
  \bibinfo {author} {\bibfnamefont {C.}~\bibnamefont {Gies}}, \bibinfo {author}
  {\bibfnamefont {P.}~\bibnamefont {Gartner}},\ and\ \bibinfo {author}
  {\bibfnamefont {F.}~\bibnamefont {Jahnke}},\ }\href
  {https://doi.org/10.1364/OE.20.005335} {\bibfield  {journal} {\bibinfo
  {journal} {Opt. Express}\ }\textbf {\bibinfo {volume} {20}},\ \bibinfo
  {pages} {5335} (\bibinfo {year} {2012})}\BibitemShut {NoStop}%
\bibitem [{\citenamefont {Kim}\ \emph {et~al.}(2018)\citenamefont {Kim},
  \citenamefont {Aghaeimeibodi}, \citenamefont {Richardson}, \citenamefont
  {Leavitt},\ and\ \citenamefont {Waks}}]{Kim2018}%
  \BibitemOpen
  \bibfield  {author} {\bibinfo {author} {\bibfnamefont {J.-H.}\ \bibnamefont
  {Kim}}, \bibinfo {author} {\bibfnamefont {S.}~\bibnamefont {Aghaeimeibodi}},
  \bibinfo {author} {\bibfnamefont {C.~J.~K.}\ \bibnamefont {Richardson}},
  \bibinfo {author} {\bibfnamefont {R.~P.}\ \bibnamefont {Leavitt}},\ and\
  \bibinfo {author} {\bibfnamefont {E.}~\bibnamefont {Waks}},\ }\href
  {https://doi.org/10.1021/acs.nanolett.8b01133} {\bibfield  {journal}
  {\bibinfo  {journal} {Nano Letters}\ }\textbf {\bibinfo {volume} {18}},\
  \bibinfo {pages} {4734} (\bibinfo {year} {2018})},\ \bibinfo {note} {pMID:
  29966093}\BibitemShut {NoStop}%
\bibitem [{\citenamefont {Fischer}\ \emph {et~al.}(2017)\citenamefont
  {Fischer}, \citenamefont {Hanschke}, \citenamefont {Wierzbowski},
  \citenamefont {Simmet}, \citenamefont {Dory}, \citenamefont {Finley},
  \citenamefont {Vu\v{c}kovi\'{c}},\ and\ \citenamefont
  {M\"uller}}]{Fischer2017}%
  \BibitemOpen
  \bibfield  {author} {\bibinfo {author} {\bibfnamefont {K.}~\bibnamefont
  {Fischer}}, \bibinfo {author} {\bibfnamefont {L.}~\bibnamefont {Hanschke}},
  \bibinfo {author} {\bibfnamefont {J.}~\bibnamefont {Wierzbowski}}, \bibinfo
  {author} {\bibfnamefont {T.}~\bibnamefont {Simmet}}, \bibinfo {author}
  {\bibfnamefont {C.}~\bibnamefont {Dory}}, \bibinfo {author} {\bibfnamefont
  {J.}~\bibnamefont {Finley}}, \bibinfo {author} {\bibfnamefont
  {J.}~\bibnamefont {Vu\v{c}kovi\'{c}}},\ and\ \bibinfo {author} {\bibfnamefont
  {K.}~\bibnamefont {M\"uller}},\ }\href {https://doi.org/10.1038/nphys4052}
  {\bibfield  {journal} {\bibinfo  {journal} {Nature Physics}\ }\textbf
  {\bibinfo {volume} {13}},\ \bibinfo {pages} {649} (\bibinfo {year}
  {2017})}\BibitemShut {NoStop}%
\bibitem [{\citenamefont {Loredo}\ \emph {et~al.}(2019)\citenamefont {Loredo},
  \citenamefont {Ant\'{o}n}, \citenamefont {Reznychenko}, \citenamefont
  {Hilaire}, \citenamefont {Harouri}, \citenamefont {Millet}, \citenamefont
  {Ollivier}, \citenamefont {Somaschi}, \citenamefont {De~Santis},
  \citenamefont {Lema\^{i}tre}, \citenamefont {Sagnes}, \citenamefont {Lanco},
  \citenamefont {Auff\`{e}ves}, \citenamefont {Krebs},\ and\ \citenamefont
  {Senellart}}]{Loredo2019}%
  \BibitemOpen
  \bibfield  {author} {\bibinfo {author} {\bibfnamefont {J.~C.}\ \bibnamefont
  {Loredo}}, \bibinfo {author} {\bibfnamefont {C.}~\bibnamefont {Ant\'{o}n}},
  \bibinfo {author} {\bibfnamefont {B.}~\bibnamefont {Reznychenko}}, \bibinfo
  {author} {\bibfnamefont {P.}~\bibnamefont {Hilaire}}, \bibinfo {author}
  {\bibfnamefont {A.}~\bibnamefont {Harouri}}, \bibinfo {author} {\bibfnamefont
  {C.}~\bibnamefont {Millet}}, \bibinfo {author} {\bibfnamefont
  {H.}~\bibnamefont {Ollivier}}, \bibinfo {author} {\bibfnamefont
  {N.}~\bibnamefont {Somaschi}}, \bibinfo {author} {\bibfnamefont
  {L.}~\bibnamefont {De~Santis}}, \bibinfo {author} {\bibfnamefont
  {A.}~\bibnamefont {Lema\^{i}tre}}, \bibinfo {author} {\bibfnamefont
  {I.}~\bibnamefont {Sagnes}}, \bibinfo {author} {\bibfnamefont
  {L.}~\bibnamefont {Lanco}}, \bibinfo {author} {\bibfnamefont
  {A.}~\bibnamefont {Auff\`{e}ves}}, \bibinfo {author} {\bibfnamefont
  {O.}~\bibnamefont {Krebs}},\ and\ \bibinfo {author} {\bibfnamefont
  {P.}~\bibnamefont {Senellart}},\ }\href@noop {} {\bibfield  {journal}
  {\bibinfo  {journal} {Nature Photonics}\ }\textbf {\bibinfo {volume} {13}},\
  \bibinfo {pages} {803} (\bibinfo {year} {2019})}\BibitemShut {NoStop}%
\bibitem [{\citenamefont {Hong}\ \emph {et~al.}(1987)\citenamefont {Hong},
  \citenamefont {Ou},\ and\ \citenamefont {Mandel}}]{Hong.Ou.ea:1987}%
  \BibitemOpen
  \bibfield  {author} {\bibinfo {author} {\bibfnamefont {C.~K.}\ \bibnamefont
  {Hong}}, \bibinfo {author} {\bibfnamefont {Z.~Y.}\ \bibnamefont {Ou}},\ and\
  \bibinfo {author} {\bibfnamefont {L.}~\bibnamefont {Mandel}},\ }\href
  {https://doi.org/10.1103/PhysRevLett.59.2044} {\bibfield  {journal} {\bibinfo
   {journal} {Phys. Rev. Lett.}\ }\textbf {\bibinfo {volume} {59}},\ \bibinfo
  {pages} {2044} (\bibinfo {year} {1987})}\BibitemShut {NoStop}%
\bibitem [{\citenamefont {Bennett}\ \emph {et~al.}(2016)\citenamefont
  {Bennett}, \citenamefont {Lee}, \citenamefont {Ellis}, \citenamefont {Meany},
  \citenamefont {Murray}, \citenamefont {Floether}, \citenamefont {Griffths},
  \citenamefont {Farrer}, \citenamefont {Ritchie},\ and\ \citenamefont
  {Shields}}]{Bennett2016}%
  \BibitemOpen
  \bibfield  {author} {\bibinfo {author} {\bibfnamefont {A.~J.}\ \bibnamefont
  {Bennett}}, \bibinfo {author} {\bibfnamefont {J.~P.}\ \bibnamefont {Lee}},
  \bibinfo {author} {\bibfnamefont {D.~J.~P.}\ \bibnamefont {Ellis}}, \bibinfo
  {author} {\bibfnamefont {T.}~\bibnamefont {Meany}}, \bibinfo {author}
  {\bibfnamefont {E.}~\bibnamefont {Murray}}, \bibinfo {author} {\bibfnamefont
  {F.~F.}\ \bibnamefont {Floether}}, \bibinfo {author} {\bibfnamefont {J.~P.}\
  \bibnamefont {Griffths}}, \bibinfo {author} {\bibfnamefont {I.}~\bibnamefont
  {Farrer}}, \bibinfo {author} {\bibfnamefont {D.~A.}\ \bibnamefont
  {Ritchie}},\ and\ \bibinfo {author} {\bibfnamefont {A.~J.}\ \bibnamefont
  {Shields}},\ }\href {https://www.ncbi.nlm.nih.gov/pmc/articles/PMC4846434/}
  {\bibfield  {journal} {\bibinfo  {journal} {Science advances}\ }\textbf
  {\bibinfo {volume} {2}},\ \bibinfo {pages} {e1501256} (\bibinfo {year}
  {2016})}\BibitemShut {NoStop}%
\bibitem [{\citenamefont {Camacho}\ \emph {et~al.}(2007)\citenamefont
  {Camacho}, \citenamefont {Pack}, \citenamefont {Howell}, \citenamefont
  {Schweinsberg},\ and\ \citenamefont {Boyd}}]{Camacho.Pack.ea:2007}%
  \BibitemOpen
  \bibfield  {author} {\bibinfo {author} {\bibfnamefont {R.~M.}\ \bibnamefont
  {Camacho}}, \bibinfo {author} {\bibfnamefont {M.~V.}\ \bibnamefont {Pack}},
  \bibinfo {author} {\bibfnamefont {J.~C.}\ \bibnamefont {Howell}}, \bibinfo
  {author} {\bibfnamefont {A.}~\bibnamefont {Schweinsberg}},\ and\ \bibinfo
  {author} {\bibfnamefont {R.~W.}\ \bibnamefont {Boyd}},\ }\href
  {https://doi.org/10.1103/PhysRevLett.98.153601} {\bibfield  {journal}
  {\bibinfo  {journal} {Phys. Rev. Lett.}\ }\textbf {\bibinfo {volume} {98}},\
  \bibinfo {pages} {153601} (\bibinfo {year} {2007})}\BibitemShut {NoStop}%
\bibitem [{\citenamefont {Akopian}\ \emph {et~al.}(2011)\citenamefont
  {Akopian}, \citenamefont {Wang}, \citenamefont {Rastelli}, \citenamefont
  {Schmidt},\ and\ \citenamefont {Zwiller}}]{Akopian.Wang.ea:2011}%
  \BibitemOpen
  \bibfield  {author} {\bibinfo {author} {\bibfnamefont {N.}~\bibnamefont
  {Akopian}}, \bibinfo {author} {\bibfnamefont {L.}~\bibnamefont {Wang}},
  \bibinfo {author} {\bibfnamefont {A.}~\bibnamefont {Rastelli}}, \bibinfo
  {author} {\bibfnamefont {O.~G.}\ \bibnamefont {Schmidt}},\ and\ \bibinfo
  {author} {\bibfnamefont {V.}~\bibnamefont {Zwiller}},\ }\href
  {http://dx.doi.org/10.1038/nphoton.2011.16} {\bibfield  {journal} {\bibinfo
  {journal} {Nature Photon.}\ }\textbf {\bibinfo {volume} {5}},\ \bibinfo
  {pages} {230} (\bibinfo {year} {2011})}\BibitemShut {NoStop}%
\bibitem [{\citenamefont {Vural}\ \emph {et~al.}(2018)\citenamefont {Vural},
  \citenamefont {Portalupi}, \citenamefont {Maisch}, \citenamefont {Kern},
  \citenamefont {Weber}, \citenamefont {Jetter}, \citenamefont {Wrachtrup},
  \citenamefont {L\"{o}w}, \citenamefont {Gerhardt},\ and\ \citenamefont
  {Michler}}]{Vural.Portalupi.ea:2018}%
  \BibitemOpen
  \bibfield  {author} {\bibinfo {author} {\bibfnamefont {H.}~\bibnamefont
  {Vural}}, \bibinfo {author} {\bibfnamefont {S.~L.}\ \bibnamefont
  {Portalupi}}, \bibinfo {author} {\bibfnamefont {J.}~\bibnamefont {Maisch}},
  \bibinfo {author} {\bibfnamefont {S.}~\bibnamefont {Kern}}, \bibinfo {author}
  {\bibfnamefont {J.~H.}\ \bibnamefont {Weber}}, \bibinfo {author}
  {\bibfnamefont {M.}~\bibnamefont {Jetter}}, \bibinfo {author} {\bibfnamefont
  {J.}~\bibnamefont {Wrachtrup}}, \bibinfo {author} {\bibfnamefont
  {R.}~\bibnamefont {L\"{o}w}}, \bibinfo {author} {\bibfnamefont
  {I.}~\bibnamefont {Gerhardt}},\ and\ \bibinfo {author} {\bibfnamefont
  {P.}~\bibnamefont {Michler}},\ }\href
  {https://doi.org/10.1364/OPTICA.5.000367} {\bibfield  {journal} {\bibinfo
  {journal} {Optica}\ }\textbf {\bibinfo {volume} {5}},\ \bibinfo {pages} {367}
  (\bibinfo {year} {2018})}\BibitemShut {NoStop}%
\bibitem [{\citenamefont {Trotta}\ \emph {et~al.}(2016)\citenamefont {Trotta},
  \citenamefont {Mart\'in-S\'anchez}, \citenamefont {Wildmann}, \citenamefont
  {Piredda}, \citenamefont {Reindl}, \citenamefont {Schimpf}, \citenamefont
  {Zallo}, \citenamefont {Stroj}, \citenamefont {Edlinger},\ and\ \citenamefont
  {Rastelli}}]{Trotta.Martin-Sanchez.ea:2016}%
  \BibitemOpen
  \bibfield  {author} {\bibinfo {author} {\bibfnamefont {R.}~\bibnamefont
  {Trotta}}, \bibinfo {author} {\bibfnamefont {J.}~\bibnamefont
  {Mart\'in-S\'anchez}}, \bibinfo {author} {\bibfnamefont {J.~S.}\ \bibnamefont
  {Wildmann}}, \bibinfo {author} {\bibfnamefont {G.}~\bibnamefont {Piredda}},
  \bibinfo {author} {\bibfnamefont {M.}~\bibnamefont {Reindl}}, \bibinfo
  {author} {\bibfnamefont {C.}~\bibnamefont {Schimpf}}, \bibinfo {author}
  {\bibfnamefont {E.}~\bibnamefont {Zallo}}, \bibinfo {author} {\bibfnamefont
  {S.}~\bibnamefont {Stroj}}, \bibinfo {author} {\bibfnamefont
  {J.}~\bibnamefont {Edlinger}},\ and\ \bibinfo {author} {\bibfnamefont
  {A.}~\bibnamefont {Rastelli}},\ }\href
  {http://dx.doi.org/10.1038/ncomms10375} {\bibfield  {journal} {\bibinfo
  {journal} {Nature Comm.}\ }\textbf {\bibinfo {volume} {7}},\ \bibinfo {pages}
  {10375} (\bibinfo {year} {2016})}\BibitemShut {NoStop}%
\bibitem [{\citenamefont {Kroh}\ \emph {et~al.}(2019)\citenamefont {Kroh},
  \citenamefont {Wolters}, \citenamefont {Ahlrichs}, \citenamefont {Schell},
  \citenamefont {Thoma}, \citenamefont {Reitzenstein}, \citenamefont
  {Wildmann}, \citenamefont {Zallo}, \citenamefont {Trotta}, \citenamefont
  {Rastelli}, \citenamefont {Schmidt},\ and\ \citenamefont
  {Benson}}]{Kroh2019}%
  \BibitemOpen
  \bibfield  {author} {\bibinfo {author} {\bibfnamefont {T.}~\bibnamefont
  {Kroh}}, \bibinfo {author} {\bibfnamefont {J.}~\bibnamefont {Wolters}},
  \bibinfo {author} {\bibfnamefont {A.}~\bibnamefont {Ahlrichs}}, \bibinfo
  {author} {\bibfnamefont {A.~W.}\ \bibnamefont {Schell}}, \bibinfo {author}
  {\bibfnamefont {A.}~\bibnamefont {Thoma}}, \bibinfo {author} {\bibfnamefont
  {S.}~\bibnamefont {Reitzenstein}}, \bibinfo {author} {\bibfnamefont {J.~S.}\
  \bibnamefont {Wildmann}}, \bibinfo {author} {\bibfnamefont {E.}~\bibnamefont
  {Zallo}}, \bibinfo {author} {\bibfnamefont {R.}~\bibnamefont {Trotta}},
  \bibinfo {author} {\bibfnamefont {A.}~\bibnamefont {Rastelli}}, \bibinfo
  {author} {\bibfnamefont {O.~G.}\ \bibnamefont {Schmidt}},\ and\ \bibinfo
  {author} {\bibfnamefont {O.}~\bibnamefont {Benson}},\ }\href
  {https://doi.org/10.1038/s41598-019-50062-x} {\bibfield  {journal} {\bibinfo
  {journal} {Scientific Reports}\ }\textbf {\bibinfo {volume} {9}},\ \bibinfo
  {pages} {13728} (\bibinfo {year} {2019})}\BibitemShut {NoStop}%
\bibitem [{\citenamefont {Maisch}\ \emph {et~al.}(2020)\citenamefont {Maisch},
  \citenamefont {Vural}, \citenamefont {Jetter}, \citenamefont {Michler},
  \citenamefont {Gerhardt},\ and\ \citenamefont {Portalupi}}]{Maisch2020}%
  \BibitemOpen
  \bibfield  {author} {\bibinfo {author} {\bibfnamefont {J.}~\bibnamefont
  {Maisch}}, \bibinfo {author} {\bibfnamefont {H.}~\bibnamefont {Vural}},
  \bibinfo {author} {\bibfnamefont {M.}~\bibnamefont {Jetter}}, \bibinfo
  {author} {\bibfnamefont {P.}~\bibnamefont {Michler}}, \bibinfo {author}
  {\bibfnamefont {I.}~\bibnamefont {Gerhardt}},\ and\ \bibinfo {author}
  {\bibfnamefont {S.~L.}\ \bibnamefont {Portalupi}},\ }\href
  {https://doi.org/10.1002/qute.201900057} {\bibfield  {journal} {\bibinfo
  {journal} {Advanced Quantum Technologies}\ }\textbf {\bibinfo {volume} {3}},\
  \bibinfo {pages} {1900057} (\bibinfo {year} {2020})}\BibitemShut {NoStop}%
\bibitem [{\citenamefont {Portalupi}\ \emph {et~al.}(2016)\citenamefont
  {Portalupi}, \citenamefont {Widmann}, \citenamefont {Nawrath}, \citenamefont
  {Jetter}, \citenamefont {Michler}, \citenamefont {Wrachtrup},\ and\
  \citenamefont {Gerhardt}}]{Portalupi.Widmann.ea:2016}%
  \BibitemOpen
  \bibfield  {author} {\bibinfo {author} {\bibfnamefont {S.~L.}\ \bibnamefont
  {Portalupi}}, \bibinfo {author} {\bibfnamefont {M.}~\bibnamefont {Widmann}},
  \bibinfo {author} {\bibfnamefont {C.}~\bibnamefont {Nawrath}}, \bibinfo
  {author} {\bibfnamefont {M.}~\bibnamefont {Jetter}}, \bibinfo {author}
  {\bibfnamefont {P.}~\bibnamefont {Michler}}, \bibinfo {author} {\bibfnamefont
  {J.}~\bibnamefont {Wrachtrup}},\ and\ \bibinfo {author} {\bibfnamefont
  {I.}~\bibnamefont {Gerhardt}},\ }\href
  {http://dx.doi.org/10.1038/ncomms13632} {\bibfield  {journal} {\bibinfo
  {journal} {Nature Comm.}\ }\textbf {\bibinfo {volume} {7}},\ \bibinfo {pages}
  {13632} (\bibinfo {year} {2016})}\BibitemShut {NoStop}%
\bibitem [{\citenamefont {Vural}\ \emph
  {et~al.}(2020{\natexlab{a}})\citenamefont {Vural}, \citenamefont {Maisch},
  \citenamefont {Gerhardt}, \citenamefont {Jetter}, \citenamefont {Portalupi},\
  and\ \citenamefont {Michler}}]{Vural.Maisch.ea:2020}%
  \BibitemOpen
  \bibfield  {author} {\bibinfo {author} {\bibfnamefont {H.}~\bibnamefont
  {Vural}}, \bibinfo {author} {\bibfnamefont {J.}~\bibnamefont {Maisch}},
  \bibinfo {author} {\bibfnamefont {I.}~\bibnamefont {Gerhardt}}, \bibinfo
  {author} {\bibfnamefont {M.}~\bibnamefont {Jetter}}, \bibinfo {author}
  {\bibfnamefont {S.~L.}\ \bibnamefont {Portalupi}},\ and\ \bibinfo {author}
  {\bibfnamefont {P.}~\bibnamefont {Michler}},\ }\href
  {https://doi.org/10.1103/PhysRevB.101.161401} {\bibfield  {journal} {\bibinfo
   {journal} {Phys. Rev. B}\ }\textbf {\bibinfo {volume} {101}},\ \bibinfo
  {pages} {161401} (\bibinfo {year} {2020}{\natexlab{a}})}\BibitemShut
  {NoStop}%
\bibitem [{\citenamefont {Shi}\ \emph {et~al.}(2007)\citenamefont {Shi},
  \citenamefont {Boyd}, \citenamefont {Camacho}, \citenamefont {Vudyasetu},\
  and\ \citenamefont {Howell}}]{Shi.Boyd.ea:2007}%
  \BibitemOpen
  \bibfield  {author} {\bibinfo {author} {\bibfnamefont {Z.}~\bibnamefont
  {Shi}}, \bibinfo {author} {\bibfnamefont {R.~W.}\ \bibnamefont {Boyd}},
  \bibinfo {author} {\bibfnamefont {R.~M.}\ \bibnamefont {Camacho}}, \bibinfo
  {author} {\bibfnamefont {P.~K.}\ \bibnamefont {Vudyasetu}},\ and\ \bibinfo
  {author} {\bibfnamefont {J.~C.}\ \bibnamefont {Howell}},\ }\href
  {https://doi.org/10.1103/PhysRevLett.99.240801} {\bibfield  {journal}
  {\bibinfo  {journal} {Phys. Rev. Lett.}\ }\textbf {\bibinfo {volume} {99}},\
  \bibinfo {pages} {240801} (\bibinfo {year} {2007})}\BibitemShut {NoStop}%
\bibitem [{\citenamefont {Shi}\ and\ \citenamefont {Boyd}(2008)}]{Shi2008}%
  \BibitemOpen
  \bibfield  {author} {\bibinfo {author} {\bibfnamefont {Z.}~\bibnamefont
  {Shi}}\ and\ \bibinfo {author} {\bibfnamefont {R.~W.}\ \bibnamefont {Boyd}},\
  }\href@noop {} {\bibfield  {journal} {\bibinfo  {journal} {J. Opt. Soc. Am.
  B}\ }\textbf {\bibinfo {volume} {25}},\ \bibinfo {pages} {C136} (\bibinfo
  {year} {2008})}\BibitemShut {NoStop}%
\bibitem [{\citenamefont {Bortolozzo}\ \emph {et~al.}(2013)\citenamefont
  {Bortolozzo}, \citenamefont {Residori},\ and\ \citenamefont
  {Howell}}]{Bortolozzo.Residori.ea:2013}%
  \BibitemOpen
  \bibfield  {author} {\bibinfo {author} {\bibfnamefont {U.}~\bibnamefont
  {Bortolozzo}}, \bibinfo {author} {\bibfnamefont {S.}~\bibnamefont
  {Residori}},\ and\ \bibinfo {author} {\bibfnamefont {J.~C.}\ \bibnamefont
  {Howell}},\ }\href {https://doi.org/10.1364/OL.38.003107} {\bibfield
  {journal} {\bibinfo  {journal} {Opt. Lett.}\ }\textbf {\bibinfo {volume}
  {38}},\ \bibinfo {pages} {3107} (\bibinfo {year} {2013})}\BibitemShut
  {NoStop}%
\bibitem [{\citenamefont {Zentile}\ \emph {et~al.}(2015)\citenamefont
  {Zentile}, \citenamefont {Keaveney}, \citenamefont {Weller}, \citenamefont
  {Whiting}, \citenamefont {Adams},\ and\ \citenamefont
  {Hughes}}]{Zentile.Keaveney.ea:2015}%
  \BibitemOpen
  \bibfield  {author} {\bibinfo {author} {\bibfnamefont {M.~A.}\ \bibnamefont
  {Zentile}}, \bibinfo {author} {\bibfnamefont {J.}~\bibnamefont {Keaveney}},
  \bibinfo {author} {\bibfnamefont {L.}~\bibnamefont {Weller}}, \bibinfo
  {author} {\bibfnamefont {D.~J.}\ \bibnamefont {Whiting}}, \bibinfo {author}
  {\bibfnamefont {C.~S.}\ \bibnamefont {Adams}},\ and\ \bibinfo {author}
  {\bibfnamefont {I.~G.}\ \bibnamefont {Hughes}},\ }\href
  {https://doi.org/https://doi.org/10.1016/j.cpc.2014.11.023} {\bibfield
  {journal} {\bibinfo  {journal} {Computer Physics Communications}\ }\textbf
  {\bibinfo {volume} {189}},\ \bibinfo {pages} {162 } (\bibinfo {year}
  {2015})}\BibitemShut {NoStop}%
\bibitem [{\citenamefont {J\"ons}\ \emph {et~al.}(2011)\citenamefont {J\"ons},
  \citenamefont {Hafenbrak}, \citenamefont {Singh}, \citenamefont {Ding},
  \citenamefont {Plumhof}, \citenamefont {Rastelli}, \citenamefont {Schmidt},
  \citenamefont {Bester},\ and\ \citenamefont
  {Michler}}]{Joens.Hafenbrak.ea:2011}%
  \BibitemOpen
  \bibfield  {author} {\bibinfo {author} {\bibfnamefont {K.~D.}\ \bibnamefont
  {J\"ons}}, \bibinfo {author} {\bibfnamefont {R.}~\bibnamefont {Hafenbrak}},
  \bibinfo {author} {\bibfnamefont {R.}~\bibnamefont {Singh}}, \bibinfo
  {author} {\bibfnamefont {F.}~\bibnamefont {Ding}}, \bibinfo {author}
  {\bibfnamefont {J.~D.}\ \bibnamefont {Plumhof}}, \bibinfo {author}
  {\bibfnamefont {A.}~\bibnamefont {Rastelli}}, \bibinfo {author}
  {\bibfnamefont {O.~G.}\ \bibnamefont {Schmidt}}, \bibinfo {author}
  {\bibfnamefont {G.}~\bibnamefont {Bester}},\ and\ \bibinfo {author}
  {\bibfnamefont {P.}~\bibnamefont {Michler}},\ }\href
  {https://doi.org/10.1103/PhysRevLett.107.217402} {\bibfield  {journal}
  {\bibinfo  {journal} {Phys. Rev. Lett.}\ }\textbf {\bibinfo {volume} {107}},\
  \bibinfo {pages} {217402} (\bibinfo {year} {2011})}\BibitemShut {NoStop}%
\bibitem [{\citenamefont {Legero}\ \emph {et~al.}(2006)\citenamefont {Legero},
  \citenamefont {Wilk}, \citenamefont {Kuhn},\ and\ \citenamefont
  {Rempe}}]{Legero.Wilk.ea:2006}%
  \BibitemOpen
  \bibfield  {author} {\bibinfo {author} {\bibfnamefont {T.}~\bibnamefont
  {Legero}}, \bibinfo {author} {\bibfnamefont {T.}~\bibnamefont {Wilk}},
  \bibinfo {author} {\bibfnamefont {A.}~\bibnamefont {Kuhn}},\ and\ \bibinfo
  {author} {\bibfnamefont {G.}~\bibnamefont {Rempe}},\ }\href@noop {} {\emph
  {\bibinfo {title} {Characterization of Single Photons using Two-Photon
  Interference}}},\ edited by\ \bibinfo {editor} {\bibfnamefont
  {G.}~\bibnamefont {Rempe}}\ and\ \bibinfo {editor} {\bibfnamefont
  {M.}~\bibnamefont {Scully}}\ (\bibinfo  {publisher} {Elsevier},\ \bibinfo
  {address} {Amsterdam},\ \bibinfo {year} {2006})\ \bibinfo {note} {advances in
  Atomic, Molecular, and Optical Physics 53 (pp.253-289)}\BibitemShut {NoStop}%
\bibitem [{\citenamefont {Rezai}\ \emph {et~al.}(2018)\citenamefont {Rezai},
  \citenamefont {Wrachtrup},\ and\ \citenamefont {Gerhardt}}]{Rezai2018}%
  \BibitemOpen
  \bibfield  {author} {\bibinfo {author} {\bibfnamefont {M.}~\bibnamefont
  {Rezai}}, \bibinfo {author} {\bibfnamefont {J.}~\bibnamefont {Wrachtrup}},\
  and\ \bibinfo {author} {\bibfnamefont {I.}~\bibnamefont {Gerhardt}},\ }\href
  {https://doi.org/10.1103/PhysRevX.8.031026} {\bibfield  {journal} {\bibinfo
  {journal} {Phys. Rev. X}\ }\textbf {\bibinfo {volume} {8}},\ \bibinfo {pages}
  {031026} (\bibinfo {year} {2018})}\BibitemShut {NoStop}%
\bibitem [{\citenamefont {Kuhlmann}\ \emph {et~al.}(2013)\citenamefont
  {Kuhlmann}, \citenamefont {Houel}, \citenamefont {Ludwig}, \citenamefont
  {Greuter}, \citenamefont {Reuter}, \citenamefont {Wieck}, \citenamefont
  {Poggio},\ and\ \citenamefont {Warburton}}]{Kuhlmann.Houel.ea:2013}%
  \BibitemOpen
  \bibfield  {author} {\bibinfo {author} {\bibfnamefont {A.~V.}\ \bibnamefont
  {Kuhlmann}}, \bibinfo {author} {\bibfnamefont {J.}~\bibnamefont {Houel}},
  \bibinfo {author} {\bibfnamefont {A.}~\bibnamefont {Ludwig}}, \bibinfo
  {author} {\bibfnamefont {L.}~\bibnamefont {Greuter}}, \bibinfo {author}
  {\bibfnamefont {D.}~\bibnamefont {Reuter}}, \bibinfo {author} {\bibfnamefont
  {A.~D.}\ \bibnamefont {Wieck}}, \bibinfo {author} {\bibfnamefont
  {M.}~\bibnamefont {Poggio}},\ and\ \bibinfo {author} {\bibfnamefont {R.~J.}\
  \bibnamefont {Warburton}},\ }\href {http://dx.doi.org/10.1038/nphys2688}
  {\bibfield  {journal} {\bibinfo  {journal} {Nature Phys.}\ }\textbf {\bibinfo
  {volume} {9}},\ \bibinfo {pages} {570} (\bibinfo {year} {2013})}\BibitemShut
  {NoStop}%
\bibitem [{\citenamefont {Stanley}\ \emph {et~al.}(2014)\citenamefont
  {Stanley}, \citenamefont {Matthiesen}, \citenamefont {Hansom}, \citenamefont
  {Le~Gall}, \citenamefont {Schulte}, \citenamefont {Clarke},\ and\
  \citenamefont {Atat\"ure}}]{Stanley.Matthiesen.ea:2014}%
  \BibitemOpen
  \bibfield  {author} {\bibinfo {author} {\bibfnamefont {M.~J.}\ \bibnamefont
  {Stanley}}, \bibinfo {author} {\bibfnamefont {C.}~\bibnamefont {Matthiesen}},
  \bibinfo {author} {\bibfnamefont {J.}~\bibnamefont {Hansom}}, \bibinfo
  {author} {\bibfnamefont {C.}~\bibnamefont {Le~Gall}}, \bibinfo {author}
  {\bibfnamefont {C.~H.~H.}\ \bibnamefont {Schulte}}, \bibinfo {author}
  {\bibfnamefont {E.}~\bibnamefont {Clarke}},\ and\ \bibinfo {author}
  {\bibfnamefont {M.}~\bibnamefont {Atat\"ure}},\ }\href
  {https://doi.org/10.1103/PhysRevB.90.195305} {\bibfield  {journal} {\bibinfo
  {journal} {Phys. Rev. B}\ }\textbf {\bibinfo {volume} {90}},\ \bibinfo
  {pages} {195305} (\bibinfo {year} {2014})}\BibitemShut {NoStop}%
\bibitem [{\citenamefont {Schimpf}\ \emph {et~al.}(2019)\citenamefont
  {Schimpf}, \citenamefont {Reindl}, \citenamefont {Klenovsk\'{y}},
  \citenamefont {Fromherz}, \citenamefont {Silva}, \citenamefont {Hofer},
  \citenamefont {Schneider}, \citenamefont {H\"{o}fling}, \citenamefont
  {Trotta},\ and\ \citenamefont {Rastelli}}]{Schimpf.Reindl.ea:2019}%
  \BibitemOpen
  \bibfield  {author} {\bibinfo {author} {\bibfnamefont {C.}~\bibnamefont
  {Schimpf}}, \bibinfo {author} {\bibfnamefont {M.}~\bibnamefont {Reindl}},
  \bibinfo {author} {\bibfnamefont {P.}~\bibnamefont {Klenovsk\'{y}}}, \bibinfo
  {author} {\bibfnamefont {T.}~\bibnamefont {Fromherz}}, \bibinfo {author}
  {\bibfnamefont {S.~F. C.~D.}\ \bibnamefont {Silva}}, \bibinfo {author}
  {\bibfnamefont {J.}~\bibnamefont {Hofer}}, \bibinfo {author} {\bibfnamefont
  {C.}~\bibnamefont {Schneider}}, \bibinfo {author} {\bibfnamefont
  {S.}~\bibnamefont {H\"{o}fling}}, \bibinfo {author} {\bibfnamefont
  {R.}~\bibnamefont {Trotta}},\ and\ \bibinfo {author} {\bibfnamefont
  {A.}~\bibnamefont {Rastelli}},\ }\href {https://doi.org/10.1364/OE.27.035290}
  {\bibfield  {journal} {\bibinfo  {journal} {Opt. Express}\ }\textbf {\bibinfo
  {volume} {27}},\ \bibinfo {pages} {35290} (\bibinfo {year}
  {2019})}\BibitemShut {NoStop}%
\bibitem [{\citenamefont {Vural}\ \emph
  {et~al.}(2020{\natexlab{b}})\citenamefont {Vural}, \citenamefont
  {Portalupi},\ and\ \citenamefont {Michler}}]{Vural2020}%
  \BibitemOpen
  \bibfield  {author} {\bibinfo {author} {\bibfnamefont {H.}~\bibnamefont
  {Vural}}, \bibinfo {author} {\bibfnamefont {S.~L.}\ \bibnamefont
  {Portalupi}},\ and\ \bibinfo {author} {\bibfnamefont {P.}~\bibnamefont
  {Michler}},\ }\href {https://doi.org/10.1063/5.0010782} {\bibfield  {journal}
  {\bibinfo  {journal} {Applied Physics Letters}\ }\textbf {\bibinfo {volume}
  {117}},\ \bibinfo {pages} {030501} (\bibinfo {year}
  {2020}{\natexlab{b}})}\BibitemShut {NoStop}%
\bibitem [{\citenamefont {Michler}\ \emph {et~al.}(2000)\citenamefont
  {Michler}, \citenamefont {Kiraz}, \citenamefont {Becher}, \citenamefont
  {Schoenfeld}, \citenamefont {Petroff}, \citenamefont {Zhang}, \citenamefont
  {Hu},\ and\ \citenamefont {Imamo\u{g}lu}}]{Michler.Kiraz.ea:2000}%
  \BibitemOpen
  \bibfield  {author} {\bibinfo {author} {\bibfnamefont {P.}~\bibnamefont
  {Michler}}, \bibinfo {author} {\bibfnamefont {A.}~\bibnamefont {Kiraz}},
  \bibinfo {author} {\bibfnamefont {C.}~\bibnamefont {Becher}}, \bibinfo
  {author} {\bibfnamefont {W.~V.}\ \bibnamefont {Schoenfeld}}, \bibinfo
  {author} {\bibfnamefont {P.~M.}\ \bibnamefont {Petroff}}, \bibinfo {author}
  {\bibfnamefont {L.}~\bibnamefont {Zhang}}, \bibinfo {author} {\bibfnamefont
  {E.}~\bibnamefont {Hu}},\ and\ \bibinfo {author} {\bibfnamefont
  {A.}~\bibnamefont {Imamo\u{g}lu}},\ }\href
  {https://doi.org/10.1126/science.290.5500.2282} {\bibfield  {journal}
  {\bibinfo  {journal} {Science}\ }\textbf {\bibinfo {volume} {290}},\ \bibinfo
  {pages} {2282} (\bibinfo {year} {2000})}\BibitemShut {NoStop}%
\bibitem [{\citenamefont {Santori}\ \emph {et~al.}(2002)\citenamefont
  {Santori}, \citenamefont {Fattal}, \citenamefont {Vu\v{c}kovi\'c},
  \citenamefont {Solomon},\ and\ \citenamefont
  {Yamamoto}}]{Santori.Fattal.ea:2002}%
  \BibitemOpen
  \bibfield  {author} {\bibinfo {author} {\bibfnamefont {C.}~\bibnamefont
  {Santori}}, \bibinfo {author} {\bibfnamefont {D.}~\bibnamefont {Fattal}},
  \bibinfo {author} {\bibfnamefont {J.}~\bibnamefont {Vu\v{c}kovi\'c}},
  \bibinfo {author} {\bibfnamefont {G.~S.}\ \bibnamefont {Solomon}},\ and\
  \bibinfo {author} {\bibfnamefont {Y.}~\bibnamefont {Yamamoto}},\ }\href
  {http://dx.doi.org/10.1038/nature01086} {\bibfield  {journal} {\bibinfo
  {journal} {Nature}\ }\textbf {\bibinfo {volume} {419}},\ \bibinfo {pages}
  {594} (\bibinfo {year} {2002})}\BibitemShut {NoStop}%
\bibitem [{\citenamefont {Weber}\ \emph {et~al.}(2019)\citenamefont {Weber},
  \citenamefont {Kambs}, \citenamefont {Kettler}, \citenamefont {Kern},
  \citenamefont {Maisch}, \citenamefont {Vural}, \citenamefont {Jetter},
  \citenamefont {Portalupi}, \citenamefont {Becher},\ and\ \citenamefont
  {Michler}}]{Weber.Kambs.ea:2019}%
  \BibitemOpen
  \bibfield  {author} {\bibinfo {author} {\bibfnamefont {J.~H.}\ \bibnamefont
  {Weber}}, \bibinfo {author} {\bibfnamefont {B.}~\bibnamefont {Kambs}},
  \bibinfo {author} {\bibfnamefont {J.}~\bibnamefont {Kettler}}, \bibinfo
  {author} {\bibfnamefont {S.}~\bibnamefont {Kern}}, \bibinfo {author}
  {\bibfnamefont {J.}~\bibnamefont {Maisch}}, \bibinfo {author} {\bibfnamefont
  {H.}~\bibnamefont {Vural}}, \bibinfo {author} {\bibfnamefont
  {M.}~\bibnamefont {Jetter}}, \bibinfo {author} {\bibfnamefont {S.~L.}\
  \bibnamefont {Portalupi}}, \bibinfo {author} {\bibfnamefont {C.}~\bibnamefont
  {Becher}},\ and\ \bibinfo {author} {\bibfnamefont {P.}~\bibnamefont
  {Michler}},\ }\href {https://doi.org/10.1038/s41565-018-0279-8} {\bibfield
  {journal} {\bibinfo  {journal} {Nature Nanotechnology}\ }\textbf {\bibinfo
  {volume} {14}},\ \bibinfo {pages} {23} (\bibinfo {year} {2019})}\BibitemShut
  {NoStop}%
\bibitem [{\citenamefont {Loudon}(2000)}]{Loudon2000}%
  \BibitemOpen
  \bibfield  {author} {\bibinfo {author} {\bibfnamefont {R.}~\bibnamefont
  {Loudon}},\ }\href {https://books.google.de/books?id=AEkfajgqldoC} {\emph
  {\bibinfo {title} {The Quantum Theory of Light}}}\ (\bibinfo  {publisher}
  {OUP Oxford},\ \bibinfo {year} {2000})\BibitemShut {NoStop}%
\end{thebibliography}%

\end{document}